\documentclass[conference,10pt,final,letterpaper]{IEEEtran}
\usepackage{graphics}                 
\usepackage{graphicx}
\usepackage[cmex10]{amsmath}
\usepackage{amssymb, amsfonts}
\usepackage{algorithmic}
\usepackage{algorithm}
\usepackage{cite}

\usepackage{verbatim}

\ifCLASSOPTIONcompsoc
\usepackage[tight,normalsize,sf,SF]{subfigure}
\else
\usepackage[tight,footnotesize]{subfigure}
\fi

\usepackage{subfigure}
\graphicspath{{figures/}}

\begin{document}

\title{A Cross-Layer Design Based on Geographic Information for Cooperative Wireless Networks}
\author{\IEEEauthorblockN{Teck Aguilar\IEEEauthorrefmark{1}, Mohamed Chedly Ghedira\IEEEauthorrefmark{1}, Syue-Ju Syue\IEEEauthorrefmark{2}, Vincent Gauthier\IEEEauthorrefmark{1}, Hossam Afifi\IEEEauthorrefmark{1} and Chin-Liang Wang\IEEEauthorrefmark{2}}
\IEEEauthorblockA{\IEEEauthorrefmark{1}Lab. CNRS SAMOVAR UMR 5157, Telecom Sud Paris, Evry, France}
\IEEEauthorblockA{Emails: \{teck.aguilar, mohamed-chedly.ghedira vincent.gauthier, hossam.afifi\}@it-sudparis.eu}
\IEEEauthorblockA{\IEEEauthorrefmark{2}Institute of Communications Engineering, National Tsing Hua University, Hsinchu, Taiwan}
\IEEEauthorblockA{Emails: d949619@oz.nthu.edu.tw, clwang@ee.nthu.edu.tw}}
\maketitle

\begin{abstract}
Most of geographic routing approaches in wireless ad hoc and sensor networks do not take into consideration the medium access control (MAC) and physical layers when designing a routing protocol. In this paper, we focus on a cross-layer framework design that exploits the synergies between network, MAC, and physical layers. In the proposed CoopGeo, we use a beaconless forwarding scheme where the next hop is selected through a contention process based on the geographic position of nodes.  We optimize this Network-MAC layer interaction using a cooperative relaying technique with a relay selection scheme also based on geographic information in order to improve the system performance in terms of reliability.
\end{abstract}

\section{Introduction}
Geographic routing \cite{Stojmenovic2002} is attractive operating in ad hoc networks, because of its good scalability. It routes packets based on the geographic location of the source, the next hop and the destination nodes. The packet forwarding is carried out according to a predefined routing metric, relevant to the geographic distance information in general. These protocols are described as "greedy", because every current node chooses a neighbor that is closest to the destination as its next hop. However this greedy mechanism fails once getting into a local optimum, i.e., a current node can not find a neighbor closer to the destination than itself to forward the packet. In traditional geographic routing protocols, nodes need to send beacon messages periodically to get their neighbors' positions and then execute the greedy mechanism; however, they may encounter problems due to the mobility of nodes, and even if they can adapt the frequency of sending beacons to the degree of the network mobility for keeping the node positions updated, they can still suffer from the inaccurate position problem. Therefore in the presence of high mobility, inaccurate position information can lead to a significant decrease in packet delivery rate and fast energy consumption in wireless nodes due to media access control (MAC) layer retransmissions.

Various solutions have been proposed for the routing layer like GPSR \cite{Karp2000}, GOAFR \cite{778447}, GOAFR+ \cite{872044}. They mainly deal with the local optimum problem mentioned above, but they do not treat the MAC problem under high mobility conditions. For the MAC layer, several propositions have been made, such as \cite{1031508,958512,1098929}, where some mechanisms are proposed to handle the medium access in an energy-efficient way with an isolated MAC layer vision. Besides the routing and MAC issues, wireless channel impairments such as fading and interference also make wireless transmission a challenging task.
In modern wireless systems, cooperative diversity \cite{1246003}\cite{1246004} and its derived single relay selection \cite{Mainaud2008,4570248,Syue09} techniques effectively mitigates the channel impairments. These techniques enable the node cooperation by allowing distributed radios to jointly transmit data.

Briefly, a layered approach to wireless ad hoc and sensor networks, where each layer stack is unaware of the operation of other layers, eliminates the benefits of joint optimization across protocol layers. Hence, a joint cross-layer design between the network and MAC layers on the one hand, and a node cooperation mechanism on the other, is necessary to improve the overall network performance. With this cross-layer approach, we exploit the synergies between the different layers while satisfying the network resource constraints. The main contribution behind our cross-layer design is to integrate the network, MAC and physical layers such that the network layer will take advantage of the broadcast nature of wireless transmissions to send the packet, the MAC layer will provide us the forwarding node with respect to a predefined metric and the physical layer will propose the reliability in transmission offered by the cooperative communications.

In this paper, we propose a cross-layer design framework called CoopGeo, which performs the greedy forwarding mechanism without using beacon messages. Instead, each node broadcasts the message and each receiving node competes to forward it based on its local metric. Once determining a forwarding node that wins the contention to forward the message, we eventually apply a cooperative relaying scheme with single relay selection mechanism, where the source node and relay node jointly transmit data through the wireless channels. With CoopGeo, we improve the physical layer performance in terms of reliability, we extend the progress to destination metric to take into consideration the physical environment, and finally, we apply a mechanism to get out from the local optimum, minimizing the exchange of control messages.

\section{Network model and problem statement}\label{statement}
We consider a wireless network, represented as a graph $G(V,E)$, where $V = \{v_{1},v_{2},\ldots,v_{n}\}$ is a finite set of nodes and $E = \{e_{1},e_{2},\ldots,e_{n}\}$ a finite set of links; the sink and nodes are randomly deployed in the area. Every node is aware of its location. The set of nodes source $V_{source} = \{v_{s1},v_{s2},\ldots,v_{sn}\}$ knows the destination location. In this network, $V_{source}$ is sending a set of packets to the sink node.

The first sub-problem treated in our paper is to find a subset of forwarding nodes $P_{F} = \{v_{f1},v_{f2},\ldots,v_{fn}\}$ from a source node to the sink $\in$ $G(V,E)$, where this path represents successful delivery of a packet while avoiding the local optimum areas. The second sub-problem is to find a subset $P_{R} = \{V_{r1},v_{r2},\ldots,v_{rn}\}$ of relaying nodes to optimize the wireless communication by means of the single relay selection cooperative communication technique.

We assume that each node has a single antenna operative over frequency-flat fading channels and can only either transmit or receive information at any time slot.

\section{Cross layer design: Geographic Contention Based Forwarding and Cooperative Communications}
CoopGeo is a cross-layer framework whose objective is the delivery of data packets from a source to a destination with known coordinates. CoopGeo is composed of two modules: 1) an integrated MAC/routing protocol that uses a contention based forwarding mechanism; this module solves the first subproblem stated at section \ref{statement} giving $P_{F}$ and 2) a cooperative communication scheme which solves the second subproblem $P_{R}$.

\emph{1) Joint MAC-routing protocol:}

At the beginning, all nodes are in contention based forwarding by default and switch to the recovery process only in case of a local minimum presence. When a source node wants to send some information to a destination node, the source node triggers a competition among the potential forwarding nodes called contention-based forwarding process (CBF). The routing layer gives the MAC layer the responsibility of handling this competition by setting up timers which are related to the progress towards the destination. First, in CBF the source broadcasts the data message, and its neighbors hear this message. Second, these neighbors compete with each other to get the right to forward the packet using the timer-based contention as explained in subsection \ref{CBF}.  Third, when the other nodes hear the CTF (Clear to Forward) message from the winning node, they suppress their timers from the contention procedure, and then the winning node forwards the message after after receiving a confirmation message (Select) sent by the source. The procedure is repeated until the message is delivered to the final destination.
Since this contention-based forwarding procedure may suffer from the local optimum problem: the packet may be stuck at a node that does not have a neighbor closer to the destination than itself, we use a recovery process to bypass the problem (hole or obstacle). For this purpose, we use the method proposed in \cite{Kalosha2008} which guarantees the delivery and finds correct edges of a local planar subgraph at the forwarding node without exchanging information with all its neighbors and then we apply the traditional face routing itself. The scheme is based on the select-and-protest principle, where the neighbors respond according to a contention function to form a planar subgraph and the protest phase removes the falsely selected neighbors that will not be in the planar structure. Here, the MAC routing module gives a node $v_{fi} \in P_{F}$ as next hop.

\emph{2) Cooperative relaying}

\begin{figure}[t]
  \centering
  \subfigure[Data/CTF/Sel messages exchange]{
    \label{ctlexchA}
    \includegraphics[width=0.39\textwidth]{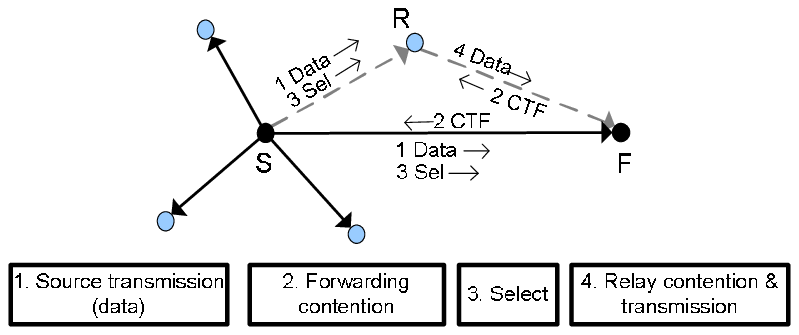}
  }
  \subfigure[Cooperative transmission]{
    \label{ctlexchB}
    \includegraphics[width=0.26\textwidth]{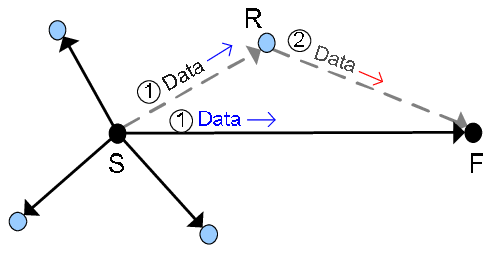}
  }
  \caption{Control message exchange and cooperative transmission}
  \label{common label}
\end{figure}

\begin{figure}[t]
  \centering
  \includegraphics[width=0.35\textwidth]{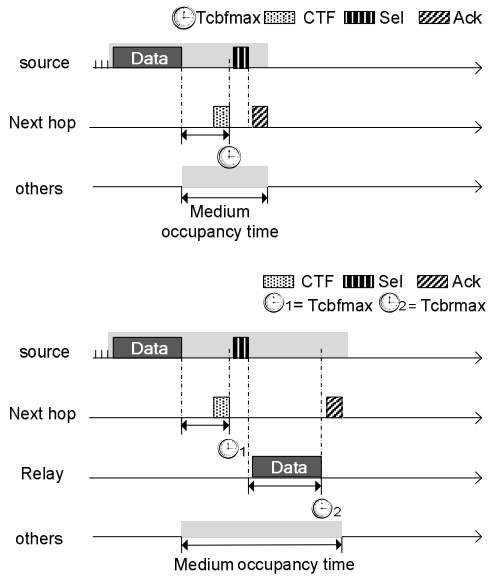}
  \caption{Cooperative and direct transmission handshaking}\label{coophand}
\end{figure}

Simultaneously to the CBF process, when the source's neighbors hear the Data/CTF/Select handshaking and the next hop node indicates that it did not decode the whole packet correctly, a single relay selection for the cooperative communication is achieved: the contention-based Relaying scheme (CBR). This means that the overhearing nodes  also compete within a time window in order to provide reliability to the data transmission process. For more details see subsection \ref{CBR}. When the Data/CTF/Select and the second data packet from the winning relay node is achieved, we have the solution to the second subproblem defined above. $v_{ri} \in P_{R}$ is used during the cooperative communication. In some cases we may have no relay nodes available. Under these circumstances, the source node will retransmit the packet using a direct transmission scheme. Fig. \ref{ctlexchA} and \ref{ctlexchB} illustrate the control packets and data exchange in cooperative transmission. For a better understanding, Fig. \ref{coophand} depicts how the direct/cooperative transmission handshaking operates.

For the cooperative transmission, the source node first sends its data and the candidate relay nodes decode the received data. At the same time, the forwarding node stores the received signal and defers the decoding for the next step. Next, the best relay forwards the previous decoded data to the forwarding node. Finally, the forwarding node jointly combines the received data sent by the source and the best relay respectively.

\subsection{Recovery process}
\begin{figure}[t]
\centerline{
\subfigure[]{\includegraphics[width=0.2\textwidth]{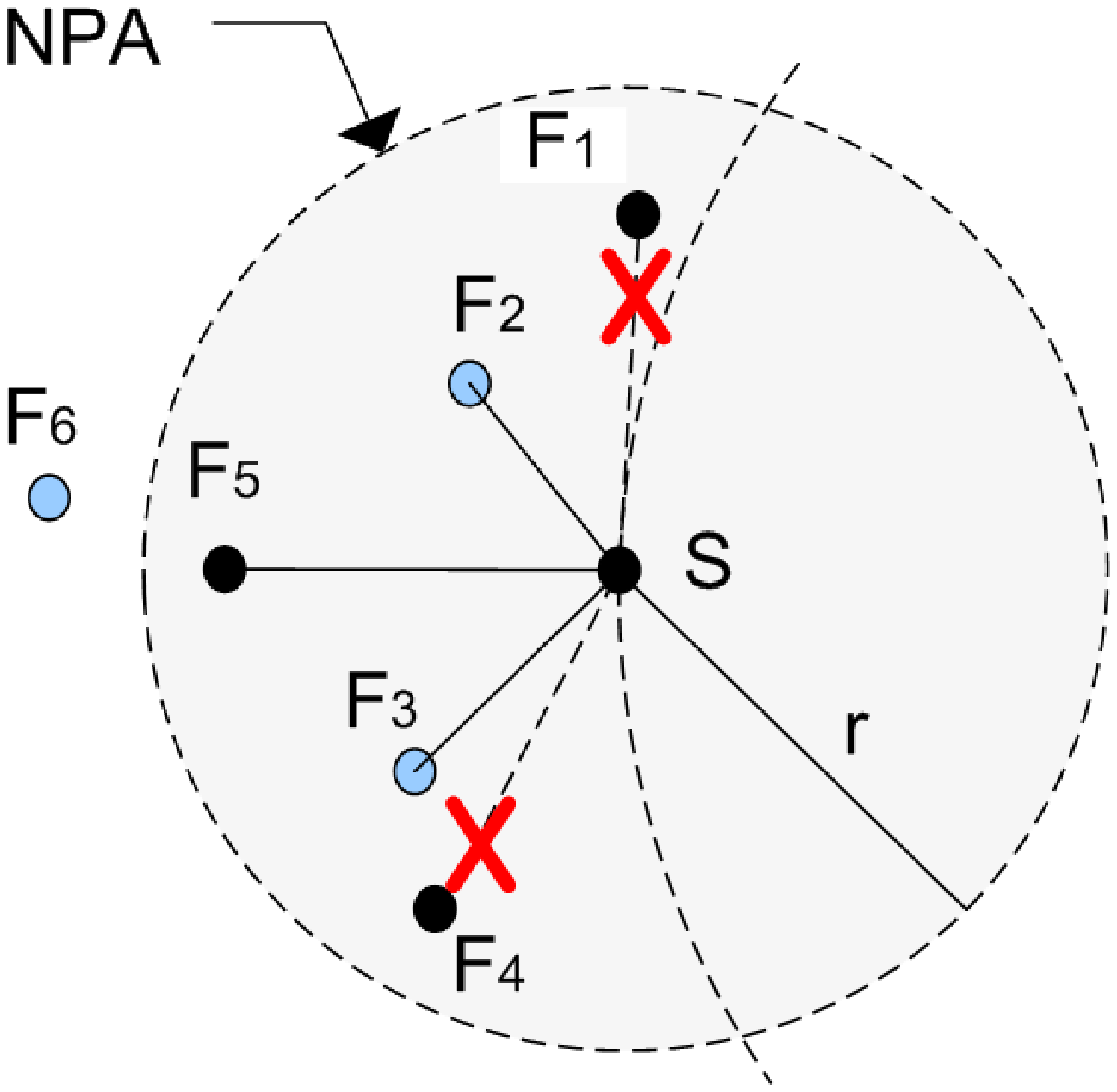} \label{rout}}
\hfil	
\subfigure[]{\includegraphics[width=0.25\textwidth]{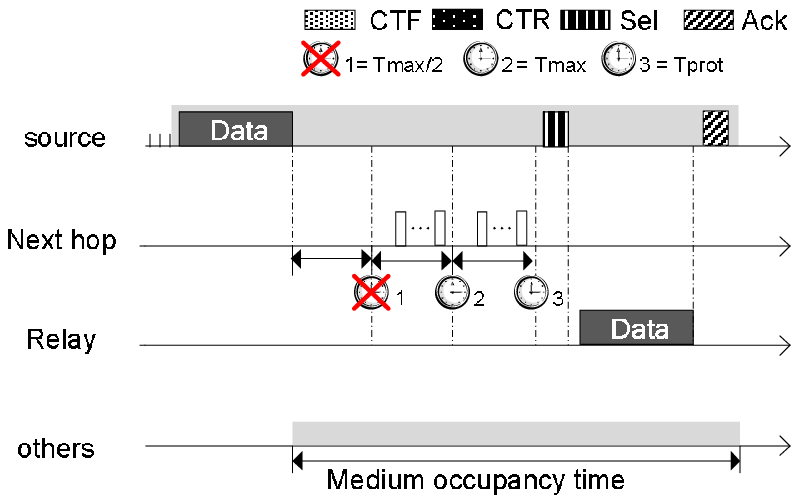} \label{recover}}}
\caption{(a) Planar subgraph (b) Chronology of messages}
\end{figure}

As we mention before, the routing process could get into a local optimum, detected when the timer of the sender reaches Tmax/2 without any reception of CTFs from the neighbors at PPA. In this case, the node switches from the contention based forwarding mode to the recovery mode. Our recovery strategy is based on the Beaconless Forwarder Planarization (BFP) proposed by Kalosha et al. in \cite{Kalosha2008}. Face traversal on a planar subgraph, is an efficient recovery method for geographic routing because it is loop free and guarantees the message delivery \cite{Bose2001}. The planar subgraph used, however, requires the knowledge of the whole neighborhood by means of beacon exchanges. In BFP, no beacon exchanges are needed. BFP, consists of two phases: the selection and the protest phase. The selection phase first aims to construct a temporal planar subgraph, protest phase is used to remove falsely selected neighbors from the temporal subgraph and get a final planar structure to apply the face traversal process.
As depicted in Fig. \ref{rout}, in the selection phase, neighbors that were not well placed to respond between 0 and Tmax/2 (located at NPA area) begin to respond by sending CTF messages after the expiration of their timers that are now set with respect the their distance to $S$ and not to $D$ as in the greedy phase. If a neighbor $F_{2}$ receives a CTF from another node $F_{1}$ and $F_{2}$ lies in the proximity region of $S-F_{1}$, then $F_{2}$ cancels its timer and remains quiet. Kalosha et al. calls this mechanism "suppression" and $F_{2}$ a hidden node. Hidden nodes listen to other nodes after their timer expiry. The hidden node $F_{3}$ remains listening to all messages even after canceling its timer, and prepares itself to protest if it receives a CTS from another neighbor i.e. $F_{3}$ lies in the proximity region of $S-F_{4}$ which is then called a violating node and its edge should disappear from the planar subgraph. The protest is made by $F_{3}$ in the second phase.
In the protest phase, a hidden node that discovers violating nodes sets a new timer using the same function defined in \cite{Sanchez2007}. At expiration time, it sends a protest message (if no other neighbor protests for the same violating node before). At the end, the sender S can build its planar subgraph according to different CTF and protest messages that it receives and apply the traditional face routing to the final subgraph.

In fig. \ref{rout} and fig. \ref{recover}, we present an example to illustrate the BFP process. Let's consider the scenario where the sender S is surrounded by six neighbors which respond in the order: $F_{1}$, $F_{4}$ and $F_{5}$ according to their timers. $F_{2}$ receives the CTF message from $F_{1}$ and becomes a hidden node, $F_{3}$ receives the CTF sent by $F_{4}$ and $F_{6}$ receives the one sent by $F_{5}$ as well. The hidden nodes are $F_{2}$, $F_{3}$ and $F_{6}$. $F_{2}$ is located in the proximity region of $F_{1}$ and $F_{3}$ in the proximity region of $F_{4}$. So, in the protest phase, $F_{2}$ protests against $F_{1}$ and $F_{3}$ protests $F_{4}$. Thus, S removes the links with violating nodes and obtains a planar subgraph that will be used to find the forwarding node. In fig \ref{recover}, we present the chronology of different responses (CTF and protest messages).

\subsection{Geographic Contention-Based Forwarder Selection}\label{CBF}
\begin{figure}[b]
  \centering
  \includegraphics[angle=90,width=0.25\textwidth]{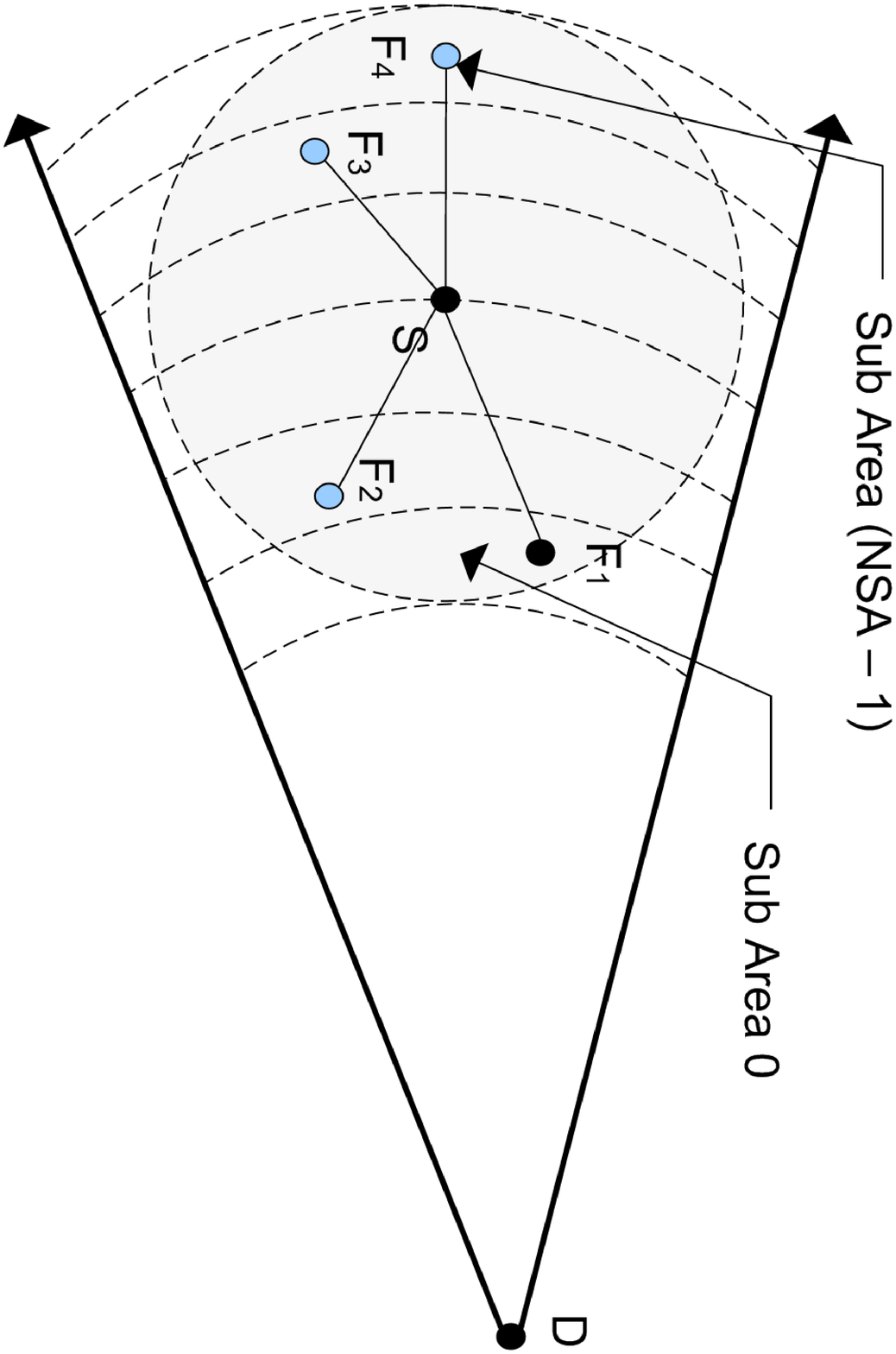}
  \caption{$F_{1}$, $F_{2}$ are in PPA area whereas $F_{3}$, $F_{4}$ are in NPA}\label{areas}
\end{figure}

To implement the CBF timers at the next hop candidate nodes called $T_{CBF}$, we use the metric proposed in \cite{Sanchez2007}. They do not only use the progress to $D$ as a criterion of goodness in the selection process but they also divided the source coverage area into a Positive Progress Area (PPA) and a Negative Progress Area (NPA) and simultaneously both areas are subdivided into groups of candidate nodes providing similar progress called Common Sub Area (CSA) so as to reduce the collision probability between candidate nodes (see Fig. \ref{areas}).

Each candidate finds out which CSA it belongs to with this formula:
\begin{equation}
CSA = \Big\lfloor NSA \times \frac{r - (d_{S,D} - d_{F_i,D})}{2r}\Big\rfloor
\end{equation}

where $NSA$ is the number of sub areas defined to divide the coverage area, $r$ the maximum progress or transmission range, so $CSA$ falls between 0 and ($NSA - 1$) where 0 corresponds to the area closest to $D$ and ($NSA -1$) to the farthest. Given the $CSA$, each candidate calculates its CBF timer according to
\begin{equation}
T_{CBF} = \Big(CSA \times \frac{T_{max}}{NSA}\Big) + rand \Big(\frac{T_{max}}{NSA}\Big)
\end{equation}

$T_{max}$ represents the maximum delay time the source node $S$ will wait for a next hop node answer and $random(x)$ obtains a random value between 0 and $x$ to reduce the collision probability. The $T_{CBF}$ function allocates the fist half of $T_{max}$ to candidates in PPA area and the other half to the rest of nodes located in NPA.

\subsection{Geographic Contention-Based Relay Selection}\label{CBR}
We proposed in \cite{Syue09} a method to select the best relay among others based on geographical information instead of relying on CSI (Channel State Information). For practical purpose the metric used to select the best available relay among other nodes will be encoded in time difference, inside a backoff-based election scheme. The election process is encoded at the CTF message during the DATA/CTF/SEL handshaking between the actual forwarding and the next hop nodes. It starts as soon as the forwarding node asks for the transmission of a second version of the packet from a relay node in order to succeed the packet transmission. Relay candidates will start their timers proportional to the metric used in \cite{Syue09}. Once the first timer expires among the candidate relay nodes, it sends the data packet to fires a response by sending a CTR packet and the timers of all other nodes are cancelled. This backoff-based election scheme enables us to get a quick and efficient answer to the question "which of my neighbors is the best suited to be a relay". We previously define the metric for the relay selection, which maximizes the SER (symbol error rate) as a function of the modulation scheme used (refer to \cite{Syue09} for a detailed explanation), where $A$ and $B$ are two constants depending on the modulation scheme. The best-suited relay $\mathbf{x}_{i}$, whose metric is $f(\mathbf{x}_{i})$, would then be the one closest to $\mathbf{x}^{*}$ which satisfies the equation (\ref{STimer4}) derived from (\ref{STimer3}).

\begin{equation}
m_i  \overset{\triangle}{=}  A^2 d_{S,R_i}^p + B d_{R_{i},D}^p,\qquad i=1,2,..,N,
\label{STimer1}
\end{equation}
\begin{equation}
f(\mathbf{x}_i)  =  A^2 \left\| \mathbf{x}_i - \mathbf{x}_S \right\|^p + B \left\|\mathbf{x}_i-\mathbf{x}_D \right\|^p
\label{STimer2}
\end{equation}
\begin{equation}
\text{minimize} \qquad f(\mathbf{x}_{i})  =  A^2 \left\|\mathbf{x}-\mathbf{x}_S \right\|^p + B\left\|\mathbf{x}-\mathbf{x}_D \right\|^p
\label{STimer3}
\end{equation}
\begin{equation}
\mathbf{x}^*  =  \frac{A^2 \mathbf{x}_S + B \mathbf{x}_D}{A^2+B}  ~~(\mbox{as } p=2)
\label{STimer4}
\end{equation}

We derive a mapping function $\mathcal{M}$, which scales our metric function $f$ into the interval $[0,1]$, where $\mathbf{x}_{max}$ is the point in a set:

\begin{equation}
\mathcal{M}(f(\mathbf{x}))= \frac{ f(\mathbf{x}) - f(\mathbf{x}^*) }{ f(\mathbf{x}_{\max }) - f(\mathbf{x}^*)}
\label{STimer5}
\end{equation}
Finally, as for the CBF timers, we use the following equation to allocate the time to each node in the contention-based Relay selection scheme (CBR).
\begin{equation}\label{Timer}
T_{CBR} = T_{max} \ \mathcal{M}(f(\mathbf{x}))  + rand \Big( \frac{2T_{max}}{NSA} \Big)
\end{equation}

\begin{figure}[bt]
\centerline{
\subfigure[]{
	\includegraphics[width=0.15\textwidth]{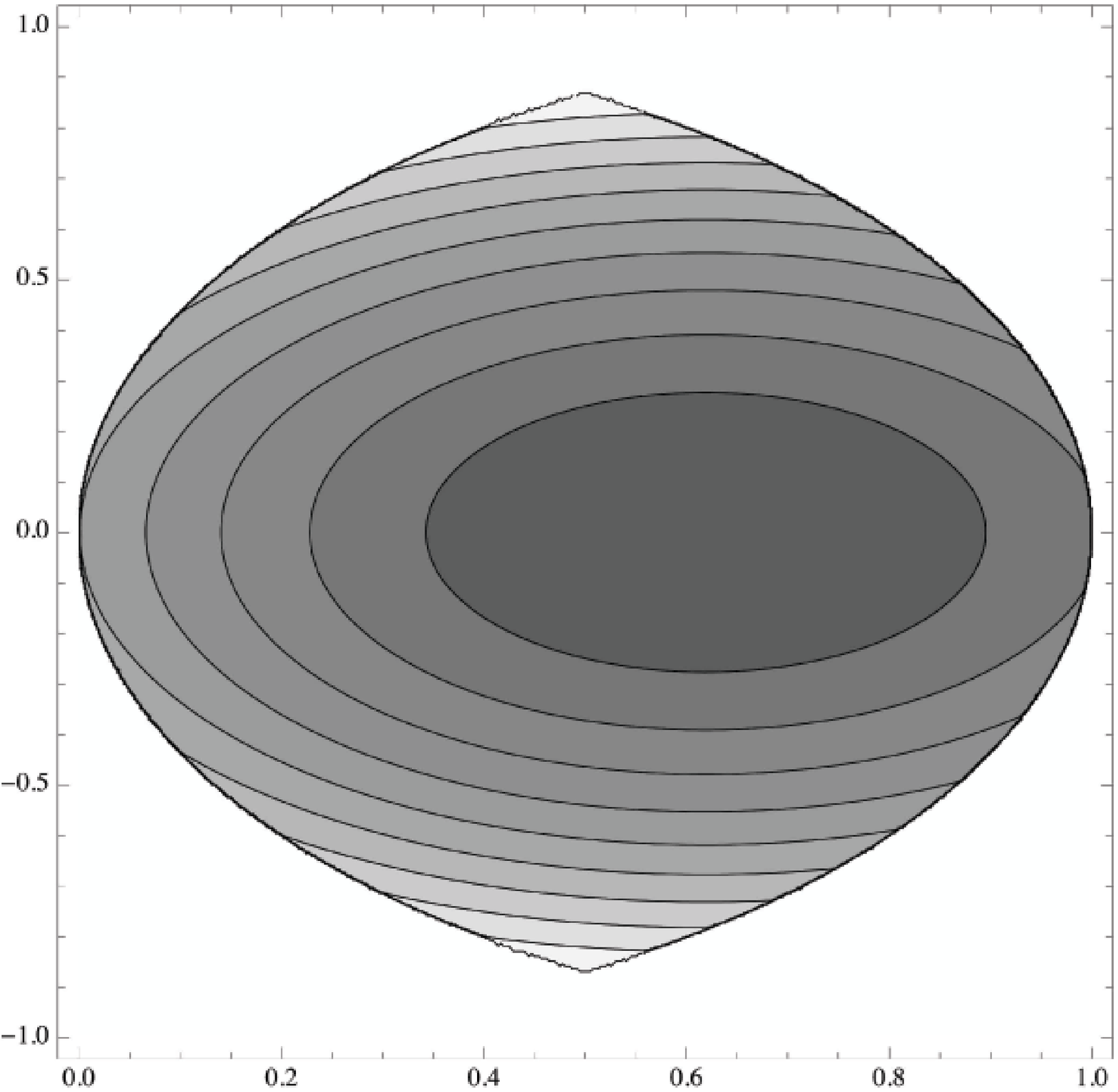}
	\label{STimerFig1a}
}
\hfil	
\subfigure[]{	
 	\includegraphics[width=0.15\textwidth]{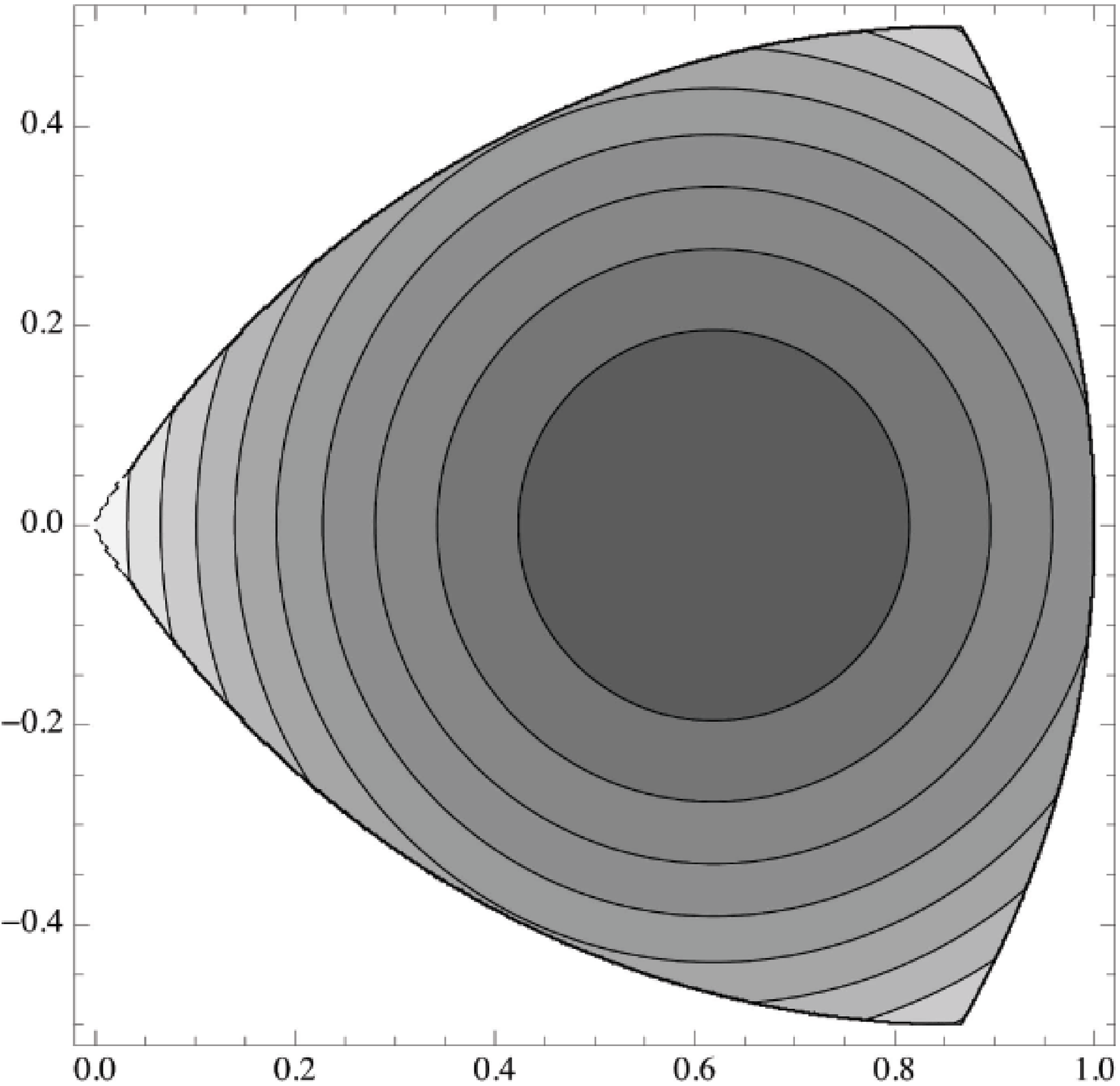}
	\label{STimerFig1b}
}}
\caption{(a) Mapping of the metric on to the set $\mathcal{C}$ (b) Mapping of the metric on to the set $\mathcal{D}$
for a normalized distance Source(0,0) Destination(1,0)}
\end{figure}

We show in Fig \ref{STimerFig1a} and \ref{STimerFig1b} the result of our metric mapped on a given set
$\mathcal{C}$. The set $\mathcal{C}$ is the union of two circles; each circle corresponds to the source node radius and the other the destination radius normalized to the unit length. Any relay $\mathbf{x}_{i} \in \mathcal{C}$ will map its metric into the set $\mathcal{C}$ like any $\mathcal{M}(f(\mathbf{x}_{i})) \in [0,1]$. In order to avoid hidden relays, our metric will be mapped onto a set $\mathcal{D}$ to be the Reuleaux triangle (Fig. \ref{STimerFig1b}).
Inside the Reuleaux triangle\cite{Fubler2003}, any Relay $\mathbf{x}_{i}$ will be at distance $R$  of any other possible relay $\mathbf{x}_{j}$, $\left\| \mathbf{x}_{i} - \mathbf{x}_{j} \right\|^{2} \leq R , \forall \mathbf{x}_{i}, \mathbf{x}_{j} \in \mathcal{D}, i \neq j$, where $R$ is the typical transmission range of a node. Using the Reuleaux
triangle as a mapping set for our metric enables us to avoid any other mechanisms in solving the problem of relays hidden from the design of the MAC layer.

\section{Performance evaluation}
In this section, we evaluate the PHY/MAC layer performance  of CoopGeo with Monte-Carlo simulations. In table \ref{table1}, we summarize the configuration settings used as input in our simulator. Our results are based on 20,000 random generated topologies where all the stations are competing to access the channel. We start by solving the two subproblems stated in section \ref{statement}, and having obtained the forwarder and relay node sets, we use them to evaluate the packet error rate, the average transmission probability and the saturated throughput.

\begin{table}[!t]
\renewcommand{\arraystretch}{1.3}
\caption{Simulation environment}
\label{table1}
\centering
\begin{tabular}{c||c||c||c}
\hline
\bfseries Input & \bfseries Value  & \bfseries Input & \bfseries Value \\
\hline\hline
Num. of neighbors & 1-20 & Tx. Power & 25 dBm \\
Channel model & Rayleigh & Average Noise & 20 dBm \\
Carrier Frequency &  2.412 Ghz & Noise Figure & 15 dBm \\
Channel Bandwidth & 22 Mhz & Packet Size & 1538 Octets \\
Modulation Type & QAM & Num. Topologies & 20000 \\
Constellation Size  & 4-64 & Simulations Run & 2000000 \\
Contention Period & 500 us & & \\
\hline
\end{tabular}
\end{table}

\begin{figure}[b]
\centerline{
\subfigure[]{
	\includegraphics[width=0.23\textwidth]{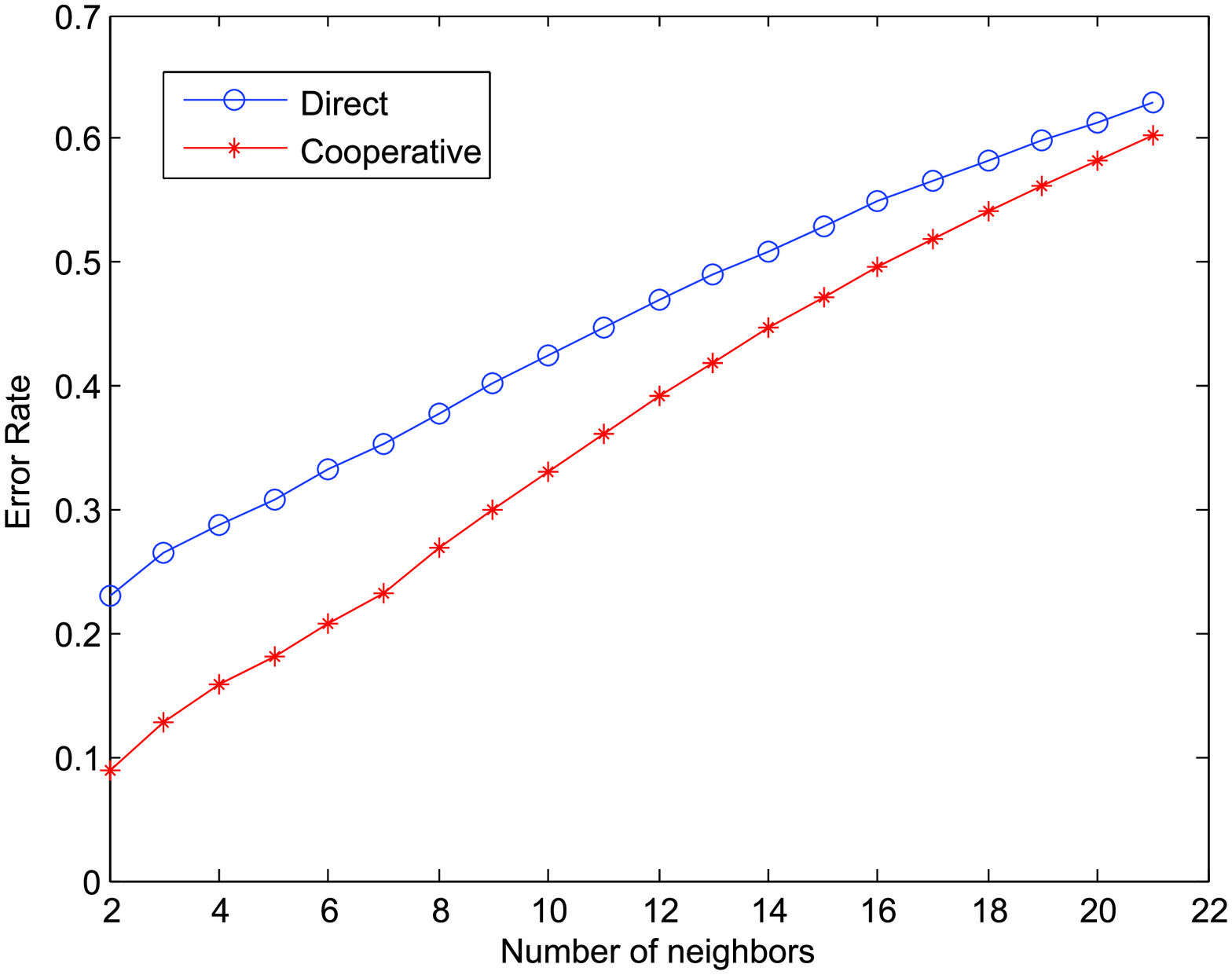}
	\label{per}
}
\hfil	
\subfigure[]{	
 	\includegraphics[width=0.23\textwidth]{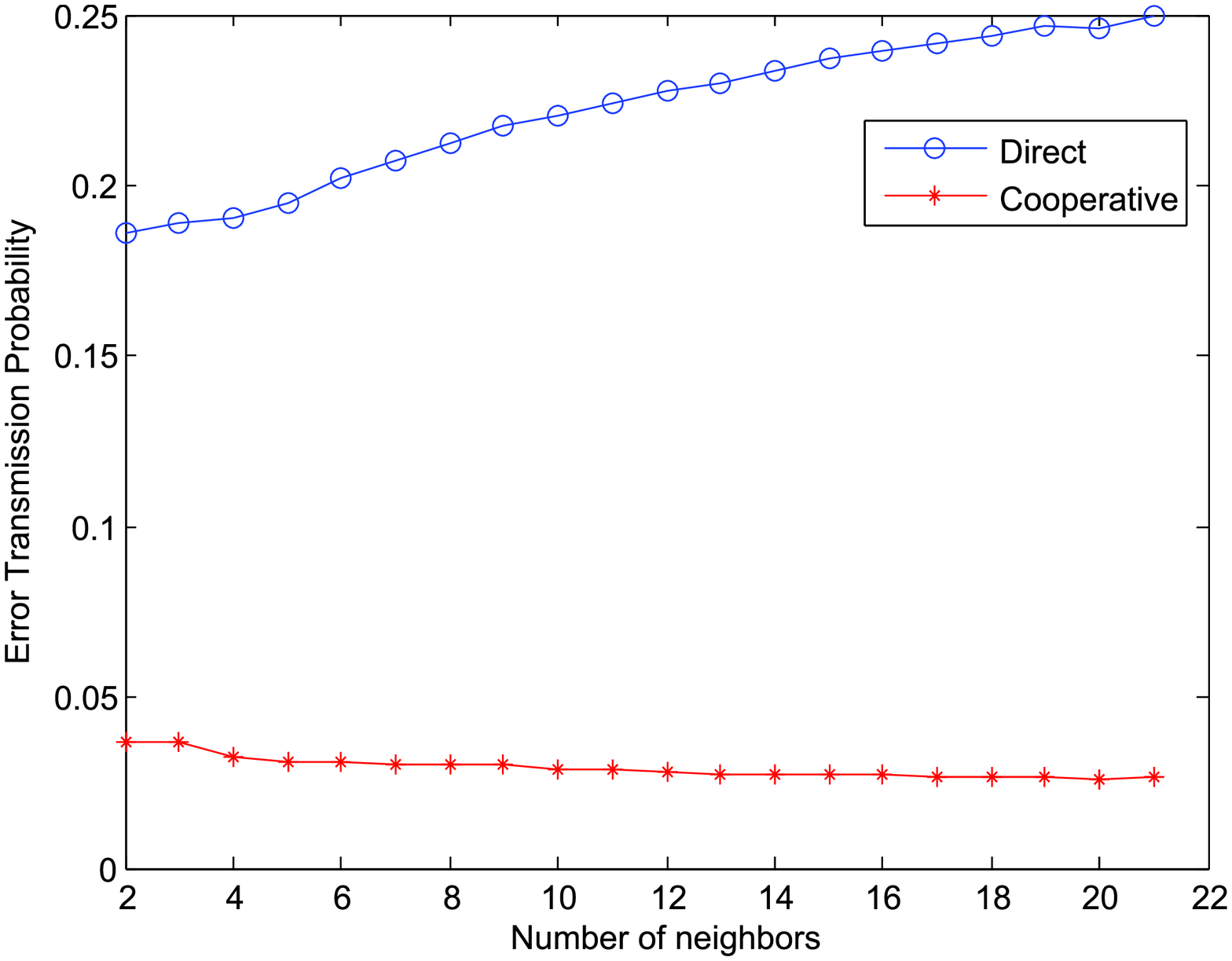}
	\label{avgtxprob}
}}
\caption{(a)Packet Error Rate (b)Transmission error probability}
\end{figure}

In Fig. \ref{per}, we show the average packet error rate of two different protocols, one is for CoopGeo using a cooperative relaying technique and the other is a BOSS\cite{Sanchez2007} like protocol without cooperative relaying. The packet error rate presented in Fig. \ref{per} includes both the probability of collision inside different contention periods and the probability of error over the wireless channel. We show that our protocol experienced a lower error rate of 2.5 times less than the traditional geographic based routing protocol in the best circumstances. We also notice that the error rate of the two protocols gets closer to each other as a function of the increased number of nodes in the neighborhood. This error rate is a function of the number of nodes and is induced by the collision probability inside the different contention periods. Furthermore, in Fig. \ref{avgtxprob}, we show that the average transmission error probability is clearly better in the cooperative case and the rate is even decreasing as the number of stations present inside the neighborhood grows. This behavior is due to the accurate selection of the relay node when more nodes are present in the neighborhood.

\begin{figure}[tb]
  \centering
  \includegraphics[scale=0.45]{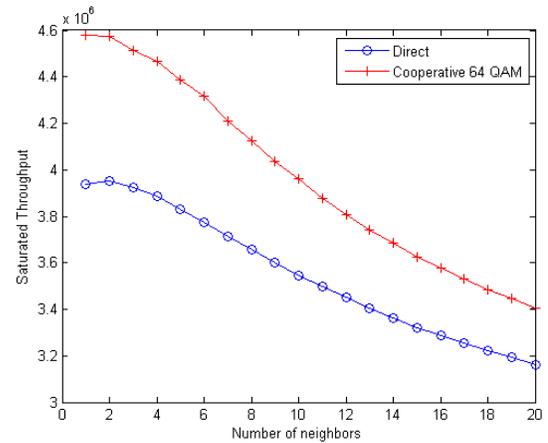}
  \caption{Saturated throughput}\label{throu}
\end{figure}

Finally, in Fig. \ref{throu}, we provide the saturated throughput of our (MAC/PHY cross-layer) CoopGeo and compare it with a traditional geographic MAC/routing approach such as BOSS. We showed that our proposal outperforms the classical scheme in terms of saturated throughput, using for this case, our framework with a 64-QAM modulation in both the source and the relay transmission channel.

\section{Conclusions and further work}
In this paper, we have proposed a cross-layer protocol, CoopGeo, based on geographic information to effectively integrate the network/MAC/physical layers for cooperative wireless networks. The proposed CoopGeo provides a joint MAC/routing protocol for forwarder selection as well as a joint MAC/physical protocol for relay selection. Simulation results demonstrate that the proposed GoopGeo can work with different densities and achieve better system performances than the existing protocol, BOSS, in terms of packet error rate, transmission error probability, and saturated throughput.
\bibliography{CoopGeo_conf}
\bibliographystyle{ieeetr}

\end{document}